\documentclass{emulateapj} \usepackage{apjfonts}

\newcommand{\mincir}{\raise
  -2.truept\hbox{\rlap{\hbox{$\sim$}}\raise5.truept \hbox{$<$}\ }}
\newcommand{\magcir}{\raise
  -2.truept\hbox{\rlap{\hbox{$\sim$}}\raise5.truept \hbox{$>$}\ }}

\shorttitle{Thermal Conduction in Simulated Galaxy Clusters}
\shortauthors{Dolag et al.}

\slugcomment{To be submitted to Astrophysical Journal Letters}

\begin{document}        

\title{Thermal conduction in simulated galaxy clusters}

\author{K.~Dolag\altaffilmark{1}, M.~Jubelgas\altaffilmark{2},
  V.~Springel\altaffilmark{2}, S. Borgani\altaffilmark{3}, E.
  Rasia\altaffilmark{1}}

\altaffiltext{1}{Dipartimento di Astronomia, Universit\`a di Padova, Padova, Italy}
\altaffiltext{2}{Max-Planck-Institut f\"ur Astrophysik, Garching, Germany}
\altaffiltext{3}{Dipartimento di Astronomia dell'Universit\`a di Trieste, Italy}

\begin{abstract}
  We study the formation of clusters of galaxies using high-resolution
  hydrodynamic cosmological simulations that include the effect of
  thermal conduction with an effective isotropic conductivity of 1/3
  the classical Spitzer value.  We find that, both for a hot ($T_{\rm
  ew}\simeq 12$ keV) and several cold ($T_{\rm ew}\simeq 2$ keV) galaxy
  clusters, the baryonic fraction converted into stars does not change
  significantly when thermal conduction is included.  However, the
  temperature profiles are modified, particularly in our simulated hot
  system, where an extended isothermal core is readily formed.  As a
  consequence of heat flowing from the inner regions of the cluster
  both to its outer parts and into its innermost resolved regions, the
  entropy profile is altered as well. This effect is almost negligible
  for the cold cluster, as expected based on the strong temperature
  dependence of the conductivity.  Our results demonstrate that while
  thermal conduction can have a significant influence on the
  properties of the intra--cluster medium of rich galaxy clusters, it
  appears unlikely to provide by itself a solution for the overcooling
  problem in clusters, or to explain the current discrepancies between
  the observed and simulated properties of the intra--cluster medium.
\end{abstract}

\keywords{conduction --- cosmology: theory --- galaxies: clusters --- methods: numerical}

\section{Introduction}

\begin{deluxetable*}{ccccccccccc}
\tablecolumns{11}
\tablewidth{0pt}
\tablecaption{Charactistics of simulated clusters}
\tablehead{\colhead{Simulation}%
& \colhead{$\left<M_{\rm vir}^{Cl1}\right>$}%
& \colhead{$M_{\rm vir}^{Cl2}$}%
& \colhead{$\left<f_{\rm cold}^{Cl1}\right>$}%
& \colhead{$f_{\rm cold}^{Cl2}$}%
& \colhead{$\left<T_{M}^{Cl1}\right>$}%
& \colhead{$T_{M}^{Cl2}$}%
& \colhead{$\left<T_{Lx}^{Cl1}\right>$}%
& \colhead{$T_{Lx}^{Cl2}$}%
& \colhead{$\left<L_x^{Cl1}\right>$}%
& \colhead{$L_x^{Cl2}$}} \startdata
csf         & 1.13$\pm0.05$ & 22.6 & 0.269$\pm1.3$ & 0.226 & 1.32$\pm0.04$ & 9.3 & 2.28$\pm0.07$ & 11.9 & 0.47$\pm0.04$ & 38.0 \\
csf + cond. & 1.08$\pm0.06$ & 22.6 & 0.261$\pm1.0$ & 0.226 & 1.30$\pm0.05$ & 9.8 & 2.15$\pm0.04$ & 12.3 & 0.43$\pm0.05$ & 54.6 
 \enddata \tablecomments{Properties of clusters when cooling, star
  formation and feedback (`csf') are included, and when conduction is
  considered as well (`csf+cond').  Columns 2--3: virial mass
  ($10^{14}\,h^{-1}M_\odot$); columns 4--5: the fraction of cold gas within
  the cluster virial regions (stars + gas below $3\times10^4\,$K); colums
  6--7: mass weighted temperature (keV); columns 8--9: emission weighted
  temperature (keV); columns 10--11: bolometric X--ray luminosity
  ($10^{44}$ erg s$^{-1}$).}
\label{tab1}
\end{deluxetable*}

\begin{figure*}[t]
\begin{center}
\resizebox{6.0cm}{!}{\includegraphics{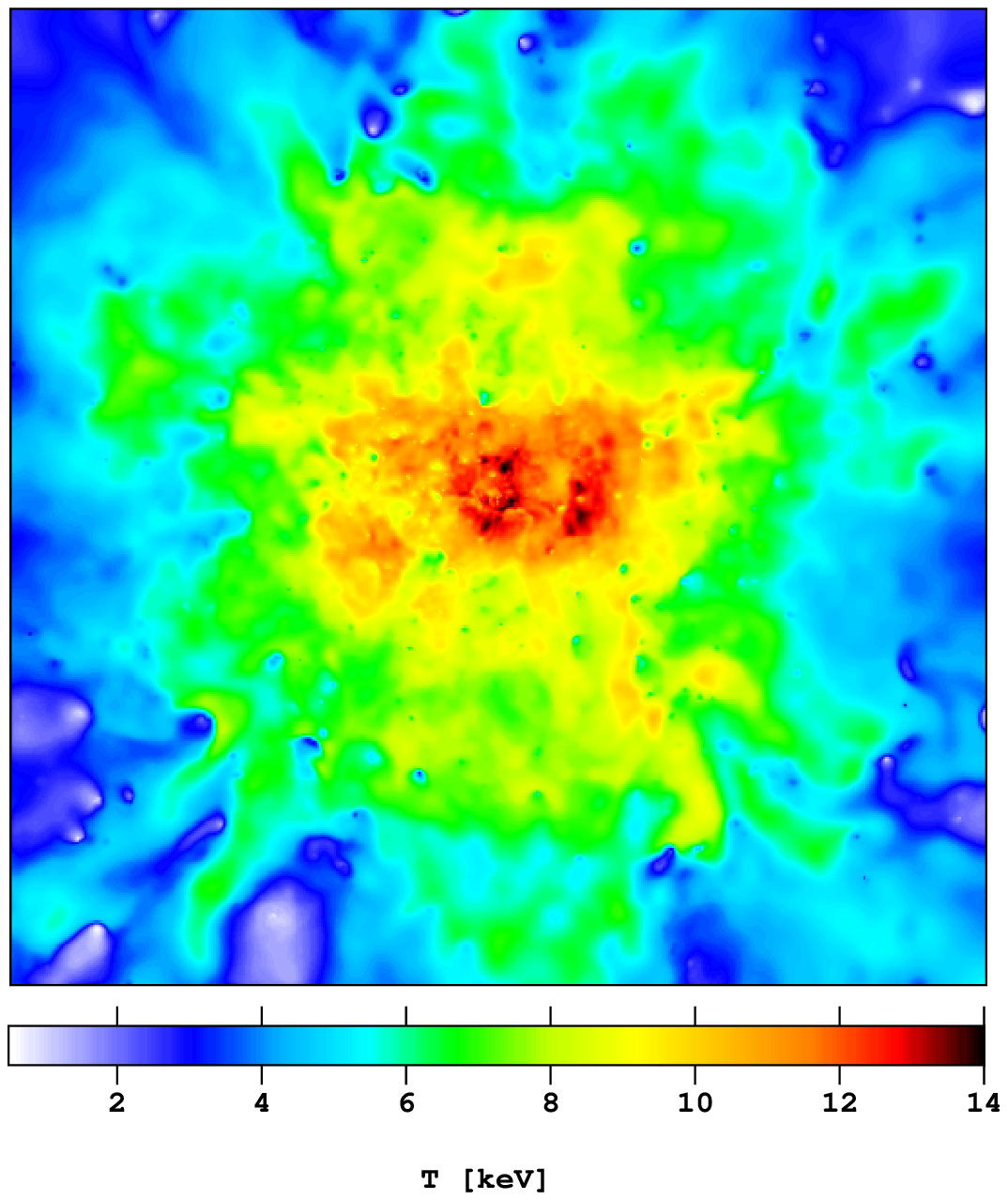}}\hspace*{1cm}%
\resizebox{6.0cm}{!}{\includegraphics{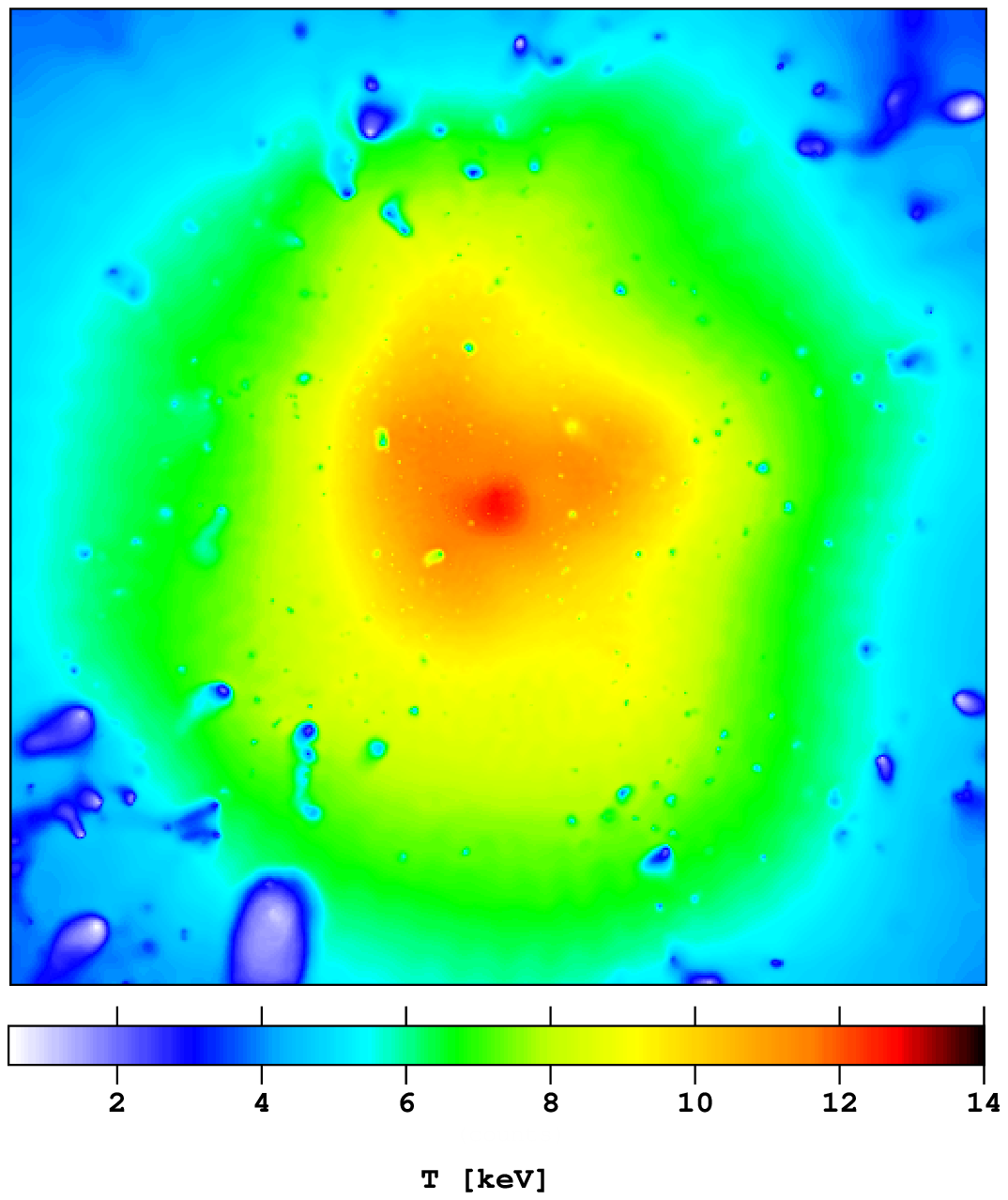}}\vspace*{-0.6cm}
\end{center}
\caption{Projected maps of mass--weighted gas temperature for our hot
  cluster (Cl2), simulated both without and with thermal conduction
  (left and right panels, respectively). Each panel shows the gas
  within a box of physical side-length of 8 Mpc on a side ($R_{\rm
  vir}\approx3.9$ Mpc), centered on the cluster center.
\label{tmap}}
\end{figure*}

Over the last few years, spatially resolved spectroscopic observations
with the XMM--Newton and Chandra satellites have provided invaluable
information about the structure of cooling gas in central cluster
regions.  Contrary to expectations based on the standard cooling--flow
model (Fabian 1994), these observations have ruled out the presence of
significant amounts of star formation and cold gas at temperatures
below 1/3rd--1/4th of the cluster virial temperature (e.g., Peterson
et al.  2001; Molendi \& Pizzolato 2001; B\"ohringer et al. 2002).
The spectroscopically measured mass--deposition rates are $\sim 10$
times smaller than those inferred from the spikes of X-ray emissivity
seen in relaxed clusters (e.g., McNamara et al. 2001; David et al.
2001).  Furthermore, measurements of temperature profiles for relaxed
hot clusters ($T\magcir 3$ keV) show that they follow an approximately
universal profile: gas is almost isothermal on scales below
one--fourth of the virial radius (De Grandi \& Molendi 2002; Pratt \&
Arnaud 2002), with a smooth decline of temperature towards the
innermost regions (e.g., Allen, Schmidt \& Fabian 2001; Johnstone et
al.  2002; Ettori et al. 2002).  These results consistently indicate
that some heating mechanism operates in cluster cores, supplying
sufficient energy to the gas to prevent it from cooling to low
($\mincir 1$ keV) temperatures.

Direct hydrodynamical simulations of cluster formation have so far
failed to reproduce these features. In particular, simulations that
include cooling and star formation find an increase of the gas
temperature in the central regions (e.g., Katz \& White 1993); here
central gas {\em does} cool out of the ICM and loses its pressure
support, so that gas flows toward the center, undergoing compressional
heating. This leads to the counterintuitive result that cooling
generates a steepening of the central temperature profiles, unlike
observed.

Narayan \& Medvedev (2001) have suggested thermal conduction as a
possible heating mechanism for the cores of galaxy clusters. This
process could transport thermal energy from the outer cluster regions
to the (slightly cooler) central gas, thereby largely offsetting its
cooling losses and stabilizing the ICM (cf. also Soker 2003). However,
conduction can only have a significant effect if the conductivity
$\kappa$ of the ICM is a sizable fraction of the Spitzer (1962) value,
$\kappa_{\rm sp}$, appropriate for an unmagnetized plasma.  In the
presence of a magnetic field, conduction is heavily suppressed
orthogonal to the field lines, so that for a tangled magnetic field,
one usually expects a relatively low, effectively isotropic
conductivity, with the amount of suppression depending on the field
topology.  However, Narayan \& Medvedev (2001) have shown that for a
chaotically tangled magnetic field, conductivities in the range
$\kappa\sim$ (0.2--0.5)$\kappa_{\rm sp}$ can be recovered. Such field
configurations may quite plausibly arise in clusters of galaxies as a
result of turbulence, so that high conductivities in some parts of the
ICM may be viable despite the presence of magnetic fields.

Using simple analytic models based on the assumption of a local
balance between radiative cooling and thermal conduction, Zakamska \&
Narayan (2003) and Voigt \& Fabian (2003) were able to reproduce the
observational data for several clusters, including their detailed
temperature profiles.  They treated the effective isotropic
conductivity $\kappa$ as a fit parameter and found good fits for
several clusters with sub--Spitzer values, while some implied
unphysically large super--Spitzer conductivities.  These results
interestingly suggest that conduction may play an important role,
while also hinting that yet another heating mechanism may be present
(cf.  also Medvedev et al. 2003).  For example, the energy feedback
from a central AGN may supplement conduction in a double heating model
(Ruszkowski \& Begelman 2002, Brighenti \& Mathews 2003).

In this Letter, we present the first cosmological hydrodynamical
simulations of cluster formation that account self-consistently for
thermal conduction, as well as radiative cooling and supernova
feedback.  Such simulations are essential to understand the highly
non-linear interplay between conduction and cooling during the
formation of clusters. In this study, we focus on the effect of
conduction on the temperature and entropy structures of clusters with
rather different temperatures of $T_{\rm Lx}\simeq 2$ and 12 keV.

\section{Numerical Simulations}

Our simulations were carried out with {\small GADGET-2}, a new version
of the parallel TreeSPH simulation code {\small GADGET} (Springel et
al.  2001).  It uses an entropy-conserving formulation of SPH
(Springel \& Hernquist 2002), and includes radiative cooling, heating
by a UV background, and a treatment of star formation and feedback
processes.  The latter is based on a sub-resolution model for the
multiphase structure of the interstellar medium (Springel \& Hernquist
2003).  We have augmented the code with a new method for treating
conduction in SPH, which is both stable and manifestly conserves
thermal energy even when individual and adaptive timesteps are used.
In our cosmological simulations, we assume an effective isotropic
conductivity parameterized as a fixed fraction of the Spitzer rate.
We also account for saturation, which can become relevant in
low-density gas.  A full discussion of our numerical implementation of
conduction is given in Jubelgas, Springel \& Dolag (2004).

We have performed simulations of galaxy clusters of two widely
differing virial mass. We refer to them as `Cl1' ($1.1 \times
10^{14}\,h^{-1}M_\odot$) and `Cl2' ($2.3\times
10^{15}\,h^{-1}M_\odot$) systems. The clusters have been extracted
from a DM--only simulation with box-size $479\,h^{-1}$Mpc of a flat
$\Lambda$CDM model with $\Omega_0=0.3$, $h=0.7$, $\sigma_8=0.9$ and
$\Omega_{\rm b}=0.04$.  Using the `Zoomed Initial Conditions'
technique (Tormen et al. 1997), we re-simulated the clusters with
higher mass and force resolution by populating their Lagrangian
regions in the initial conditions with more particles, adding
additional small-scale power appropriately.  Gas was introduced in the
high--resolution region by splitting each parent particle into a gas
and a DM particle.  The final mass--resolution of these simulations
was $m_{\rm DM}=1.13\times 10^9\,h^{-1}M_\odot$ and $m_{\rm
gas}=1.7\times 10^8\,h^{-1}M_\odot$ for dark matter and gas within the
high--resolution region, respectively.  The clusters were hence
resolved with about $4\times10^6$ and $2\times10^5$ particles,
respectively. The substantially lower computational cost of `Cl1'
systems allowed us to simulate 5 clusters within a very narrow mass
range, all yielding consistent results. The gravitational softening
length was $\epsilon=5.0\, h^{-1}$kpc (Plummer--equivalent), kept
fixed in comoving units.

For each cluster, we run simulations both with and without thermal
conduction, but we always included radiative cooling with a primordial
metallicity, and star formation. For the conduction runs, we assume a
conductivity of $\kappa = 1/3\,\kappa_{\rm sp}$, where $\kappa_{\rm
sp}\propto T^{5/2}$ is the temperature-dependent Spitzer rate for a
fully ionized, unmagnetized plasma.  Our choice for $\kappa$ is
appropriate in the presence of magnetized domains with randomly
oriented $B$--fields (e.g., Sarazin 1988), or for a chaotically
tangled magnetic field (Narayan \& Medvedev 2001).

\begin{figure}[t]
\begin{center}
\resizebox{7cm}{7cm}{\includegraphics{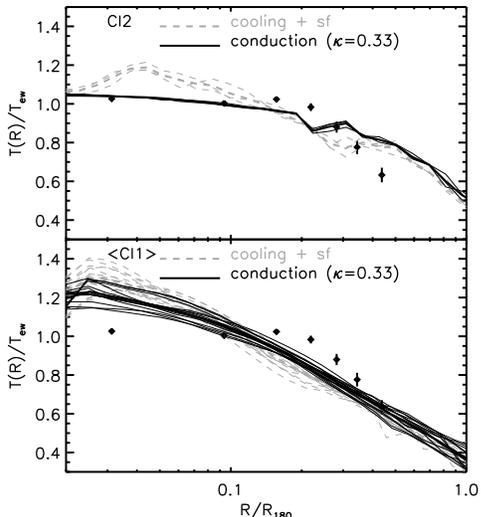}}\vspace*{-0.7cm}
\end{center}
   \caption{Comparison of projected temperature profiles for our hot
     (Cl2, upper panel) and cold clusters (Cl1, lower panel) when
     conduction is, or is not, included.  For each run, thick lines
     give the average profiles for three orthogonal projection
     directions, which are also shown individually as thin lines.  The
     bundle of lines in the lower panel illustrates the dispersion
     among our 5 simulated clusters of this mass.  For reference,
     symbols with error-bars give observational data by De Grandi \&
     Molendi (2002).
\label{tprof}}
\end{figure}

\section{Results}
An expected general effect of thermal conduction is to make the gas
more isothermal by smoothing out temperature substructure in the ICM.
This effect is clearly visible in Figure \ref{tmap}, where we compare
projected temperature maps of our massive cluster, with and without
conduction.  The cluster without conduction (left panel) shows a rich
pattern of small-scale temperature fluctuations, stemming from the
continuous stirring of the ICM by infalling galaxies. These
fluctuations are largely wiped out when conduction is included (right
panel).  Gas in large infalling galaxies can stay cooler than the
local ICM (prior to ram-pressure stripping) in the simulation without
conduction, but this same gas is conductively heated as soon as it
enters the hot cluster atmosphere in the other simulation, leading to
much more rapid thermalization.  We also note that the cluster with
conduction shows a larger, near-isothermal region near the center, and
appears somewhat hotter in its outer parts.

In Figure \ref{tprof}, we show the projected temperature profiles for
the simulated clusters, compared with observational results by
De~Grandi \& Molendi (2002) for a set of 22 clusters with $T>3\,{\rm
keV}$ observed with the Beppo--SAX satellite. For the runs that
include only cooling and star formation, we find rising temperature
profiles towards the cluster center, consistent with recent simulation
work (e.g., Lewis et al. 2000, Kay et al. 2002, Tornatore et al.
2003; Borgani et al. 2003). This behavior disagrees with observational
evidence for the presence of an isothermal regime at $R\mincir
0.2R_{180}$ (here $R_{180}$ is the radius encompassing an average
density $180\rho_{\rm crit}$) and a smooth decline in the innermost
regions (Allen, Schmidt \& Fabian 2001; De Grandi \& Molendi 2002).

However, thermal conduction significantly flattens the temperature
profiles. In fact, for the hot cluster, an isothermal core is created,
making it more similar to what is observed.  For the colder system on
the other hand, the temperature profile is almost unaffected,
consistent with expectations based on the strong dependence of the
conductivity on electron temperature, which favors thermal conduction
in hotter gas. This also implies that thermal conduction cannot easily
account for the observed self--similarity of the temperature profiles
of fairly relaxed clusters.

Interestingly, we find that the mass--weighted as well as
emission--weighted temperatures of the two simulated clusters change
by less than 10\% when conduction is included.  Conduction thus mainly
appears to re--distribute the overall thermal energy content within
the cluster, while not causing a significant heat loss to the outer
intergalactic medium as proposed by Loeb (2002).  We do however find
that the temperature of the outer parts of clusters is raised by
conduction, as seen in the temperature profiles of Fig.~\ref{tprof}.
In fact, based on our simulations, it is not clear whether conductive
heating of the innermost parts of clusters occurs in any significant
way. Instead, the dominant effect seems to be heat transport from
inner to outer parts, which can be understood as a consequence of the
falling temperature gradient obtained in simulations that only include
radiative cooling and star formation. While conduction appears to be
effective in establishing an isothermal temperature in the core, the
innermost regions subsequently do not become still cooler, which would
be required to turn around the direction of conductive heat flow and
tap the thermal reservoir at larger radii in the way proposed by
Zakamska \& Narayan (2003).  We do thus not find that conduction can
really prevent the central cooling flow; it apparently only transports
the energy gained by compressional heating of inflowing gas to outer
regions of the cluster.

In fact, the inclusion of conduction may even make the central cooling
flow stronger, depending on the resulting density and temperature
structure of the inner parts in dynamical equilibrium. For our hot
cluster, this actually seems to be the case, judging from the
bolometric X--ray luminosity, which is increased by about 40\% at
$z=0$ when conduction is included. The colder clusters on the other
hand show an essentially unchanged X-ray luminosity, consistent with
our previously found trends.

However, we note that the fraction of collapsed baryons (cold gas and
stars) in the clusters is essentially independent of conduction, as
seen in Table~\ref{tab1}, where we summarize some of the main
characteristics of the simulated clusters.  This result is not really
surprising because at high redshift, when most of the star formation
in the cluster galaxies takes place, the gas temperature in the
progenitor systems is much lower than the virial temperature reached
eventually at $z=0$, and, therefore, the effect of thermal conduction
is expected to be weak.  The amount of collapsed gas thus remains at
$f_{\rm cold}\simeq 0.20$--0.25, about a factor two larger than
indicated by observations (e.g., Balogh et al. 2001; Lin, Mohr \&
Stanford 2003), suggesting that stronger feedback processes than
included in our simulations are at work in the real universe. In any
case, conduction appears unable to resolve this overcooling problem on
its own.

\begin{figure}[t]
\begin{center}
\resizebox{7cm}{5.25cm}{\includegraphics{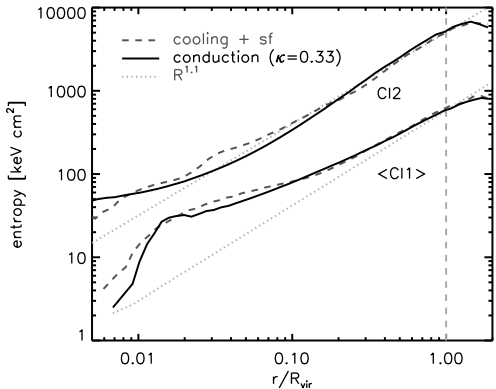}}\vspace*{-0.6cm}
\end{center}
\caption{Entropy profiles for the Cl2 cluster (upper curves) and for
  the average of the Cl1 clusters (lower curves).  The different lines
  distinguish runs with and without thermal conduction, and for pure
  gravitational heating.
\label{entr}}
\end{figure}

Another piece of information about the thermodynamical properties of
the ICM is provided by its entropy ($S = T \;n_e^{-2/3}$) profile,
which we show in Figure \ref{entr} for our cluster simulations, also
compared to results for pure gravitational heating. The effect of
cooling is that of selectively removing low entropy gas from the hot
diffuse phase in central cluster regions, such that a net entropy
increase of X--ray emitting gas compared to pure gravitational heating
simulations is seen. This has been predicted by analytic models of the
ICM (e.g. Voit et al. 2003) and has also been confirmed in direct
hydrodynamical simulations.  Interestingly, the inclusion of heat
conduction appears to reverse part of this change in the entropy
profile, but only in the hot cluster, where conduction is
efficient. Here, the entropy decreases on scales $\sim
\,$(0.01--0.1)$R_{\rm vir}$ because of conductive losses both to the
outer and innermost parts.  For the colder systems, only a very weak
modification in the region of central mass drop-out is seen. Hotter
systems hence tend to become more isentropic in central regions, which
is just the opposite of what one would observe if, instead,
pre-heating is responsible for breaking the ICM self-similarity
(Borgani et al.  2001), but the trend is consistent with a recent
analysis of entropy profiles obtained from ASCA data of galaxy groups
and clusters (Ponman et al. 2003).

\section{Conclusions}
We have presented self-consistent cosmological hydrodynamical
simulations of the formation of galaxy clusters which for the first
time included the effect of thermal conduction. In particular, we
carried out simulations of moderately poor clusters with $T_{\rm
ew}\simeq 2.0$ keV, and of a rich system with $T_{\rm ew}\simeq
12\,{\rm keV}$.  For both clusters of both masses, we compared
simulations that followed radiative cooling, star formation and
feedback with corresponding ones that also accounted for thermal
conduction with an effective isotropic conductivity of $\kappa =
\kappa_{\rm sp}/3$. Our main results can be summarized as follows:

{\bf (a)} Thermal conduction creates an isothermal core in the central
regions of our hot cluster, thus producing a temperature profile
similar to those observed. However, this effect is much less
pronounced for our poorer systems, owing to the sensitive temperature
dependence of the conductivity.  As a result, the presence of
conduction together with cooling does not lead to self--similar
temperature profiles, unlike observed for real clusters.

{\bf (b)} Compared to simulations with cooling only, conduction leads
to a small decrease of the entropy in most of the inner regions of the
hot galaxy cluster, except perhaps for the innermost part at $R\mincir
0.01\,R_{\rm vir}$. This can be understood as a result of heat flowing
from these regions both to outer parts of the cluster, and at some
level also to the innermost regions.  Again, this effect is largely
absent in the colder simulated clusters.

{\bf (c)} Conduction does not avoid the `overcooling problem'. Even
for our hot cluster, where conduction is quite efficient, we find an
essentially unchanged baryon fraction of $f_{\rm cold}\magcir 0.2$ in
cold gas and stars, which is larger than what is observed. This is
because most of the cooling and star formation takes place at high
redshift when the temperature of the diffuse gas in halos is low
enough that conduction is inefficient. Stronger feedback processes
than considered here, e.g.~energetic galactic winds, are required to
solve this problem.  We note that even for the hot cluster at $z=0$,
we do not find a temperature structure that would allow central
cooling losses to be offset by heat conduction, making it questionable
whether a detailed local balance between radiative losses and heat
conduction can arise naturally in hierarchical cluster formation.

While larger samples of simulated clusters will be required for a more
detailed assessment of the role of thermal conduction, our results
already demonstrate that conduction can have a sizeable effect on the
observational characteristics of rich galaxy clusters.  However, its
inclusion appears unlikely to to overcome the current discrepancies
between simulated and observed properties of the ICM. For instance,
conduction tends to produce different temperature profiles for cold
and hot clusters, invoking a conflict with the observed
self--similarity.  Furthermore, since conduction does not prevent
overcooling, the presence of some other heating source, perhaps AGN,
appears still required.

Admittedly, our present simulations still lack a realistic
self-consistent description of the magnetic field structure, which can
make conduction less important.  Spatial variations in the
conductivity, its interplay with gas turbulence, as well as potential
effects of anisotropic conduction due to ordered field components, can
thus not be properly taken into account.  This represents a major
uncertainty in assessing the relevance of conduction for real galaxy
clusters. It is a highly interesting task for future work to reduce
this uncertainty by a better theoretical and observational
understanding of the magnetic properties of the ICM.

\vspace*{-0.9cm}\acknowledgments The simulations were carried out on
the IBM-SP4 at CINECA, Bologna, with CPU time assigned under an
INAF-CINECA grant, and on the IBM-SP3 at Padova.  We acknowledge
useful discussions with S.~Molendi, and thank G.~Tormen for providing
the {\small ZIC} code.  K.~Dolag acknowledges support by a Marie Curie
Fellowship of the European Community program ``Human Potential'' under
contract number MCFI-2001-01227.

\end{document}